\documentclass[a4paper,fleqn,usenatbib]{mnras}

\usepackage[T1]{fontenc}
\usepackage{ae,aecompl}

\usepackage{graphicx}	
\usepackage{amsmath}	
\usepackage{amssymb}	

\def\submitted{MNRAS.}%

\title[Extended stellar structure around NGC\,288]{On the extended stellar 
structure around NGC\,288 }

\author[Andr\'es E. Piatti]{
Andr\'es E. Piatti$^{1,2}$\thanks{E-mail: andres@oac.unc.edu.ar}
\\
$^{1}$Consejo Nacional de Investigaciones Cient\'{\i}ficas y T\'ecnicas, Av. Rivadavia 1917, 
C1033AAJ, Buenos Aires, Argentina\\
$^{2}$Observatorio Astron\'omico, Universidad Nacional de C\'ordoba, Laprida 854, 5000, 
C\'ordoba, Argentina\\
}

\date{Accepted XXX. Received YYY; in original form ZZZ}

\pubyear{2017}

\begin{document}
\label{firstpage}
\pagerange{\pageref{firstpage}--\pageref{lastpage}}
\maketitle

\begin{abstract}
We report on observational evidence of an extra-tidal clumpy structure around NGC\,288
from an homogeneous coverage of a large area with the Pan-STARRS PS1 database.
The extra-tidal star population  has been disentangled from that of the Milky Way field 
by using a cleaning technique that successfully reproduced the stellar density, luminosity function
and colour distributions of MW field stars. We have produced the cluster stellar density 
radial profile and a stellar density map from independent approaches, from which we found results
in excellent agreement : the feature extends up to 3.5 times the cluster tidal radius. 
Previous works based on shallower photometric data sets
have speculated on the existence of several long tidal tails, similar to that found in Pal\,5.
The present outcome shows that NGC\,288 could
hardly have such tails, but favours the notion that interactions with the MW tidal field
has been a relatively inefficient process for stripping stars off the cluster. These results
point to the need of a renewed overall study of the external regions of Galactic
globular clusters (GGCs) in order to reliably characterise them. Hence, it will 
be possible to investigate whether there is  any connection between detected
tidal tails, extra-tidal stellar populations, extent diffuse halo-like structures 
 with the GGCs' dynamical histories in the Galaxy.

\end{abstract}




\section{Introduction}

A non-negligible number of Galactic globular clusters (GGCs) have shown
extended stellar structures around them \citep{carballobelloetal2012}.
For instance, an extended stellar halo surrounding the distant NGC\,5694 was discovered
by \citet{correntietal2011}, while  an unprecedented extra-tidal, azimuthally smooth, 
halo-like diffuse spatial extension of NGC\,1851 was found by \citet{olszewskietal2009}. \citet{p17c} also found a similar feature around 47\,Tuc. Other GGCs were found to be embedded in a diffuse stellar envelope extending to a radial distance 
of at least five time the nominal tidal radius, like M\,2 \citep{kuzmaetal2016}; and
long tidal tails have  been detected in the field of Pal\,5
 \citep{odenetal2003}, Pal\,14 \citep{sollimaetal2011}, Pal\,15 \citep{myeongetal2017},
and NGC\,7492 \citep{naverreteetal2017}.
As far as theoretical developments are considered, some N-body experiments have shown
that the detection of extended  envelopes around GGCs could be due to potential escapers 
\citep{kupperetal2010} or potential observational biases
\citep{bg2017}, among others.

As far as NGC\,288 is considered, \citet{leonetal2000} used photographic photometry to claim that the cluster
has very important tidal tails extending up to 350 pc from its centre.
They noticed that NGC\,288 had been previously observed by \citet{grillmairetal1995}, 
who found tidal extensions on a field 200$\arcmin$$\times$200$\arcmin$ 
smaller than theirs, but with the same spatial resolution (16$\arcmin$). 
\citet{leonetal2000} showed that their wavelet decomposition clearly reveals some wide structures missed by \citet{grillmairetal1995}, especially towards the south.
Particularly, they highlighted two main tidal tails, one running from
the cluster centre towards the south-east \citep[parallel to the cluster
proper motion vector, PA $\approx$ 51$\degr$;][]{dinescuetal1997} and another 
one towards the Galactic centre (PA $\approx$ 140$\degr$). They also mentioned that
the photometric error at $B$ $\sim$ 20.0 mag is $\sigma$($B$) $\sim$ 0.2 mag
and that Abell galaxy clusters affect their surface densities built using a 
{\it Gaussian} kernel of 16 arcmin, which is similar to the size of the cluster tidal radius
\citep{trageretal1993,grillmairetal1995,harris1996,miocchietal13}.

Later, \citet{grillmairetal2004} speculated on the existence of a long tidal tail of
8.5$\degr$ from the cluster centre towards the north-west -- that 
\citet{leonetal2000} did not detect -- beside its counterpart tail of 8.5$\degr$ to the south-east. 
They used 2MASS data that barely reaches the horizontal
branch of the cluster. According to \cite{go1997} and \citet{dinescuetal1999},
NGC\,288 has experienced disruption by tidal shocks more important than 
by internal relaxation and evaporation, so that tidal tails should be
expected as debris from those interactions with the Milky Way (MW). Multiple tidal tails
from the interaction with the MW potential have also been recently predicted
from numerical simulations \citep{hk2015}, while \citet{bg2017} have pointed out that
NGC\,288 is one of the GGCs  with the optimal detectability conditions
of tidal  tails, namely, a low remaining mass fraction $\mu$ -which is a
measure of its stage of dissolution -, and a high orbital phase $\phi$. \citet{dinescuetal1999} listed other seven GGCs 
with similar dynamical histories as NGC\,288, Pal\,5 being the only one with confirmed 
long tidal tails \citep{odenetal2001,odenetal2003,erkaletal2017}. The other six clusters 
do not show any observational 
hints for such an structure \citep{chch2010}.

In this paper we show from deep wide-field photometry that NGC\,288
exhibits a single extra-tidal cumply structure, as commonly seen in other GGCs 
\citep[e.g.][]{grillmairetal1995,chch2010,p17c}, no evidence of long tails is 
detected. Section\,2 deals with the description of the database used, while 
Sections 3 and 4 focus on independent approaches to produce the cluster stellar
density radial profile and a stellar density map of its outskirts. In Section 5
we discuss the outcomes to the light of previous results and pose further studies
that will be possible the address from upcoming releases of ongoing surveys. Finally,
Section 6 summaries the main conclusions of this work.

\section{Observational data}

\begin{figure}
	\includegraphics[width=\columnwidth]{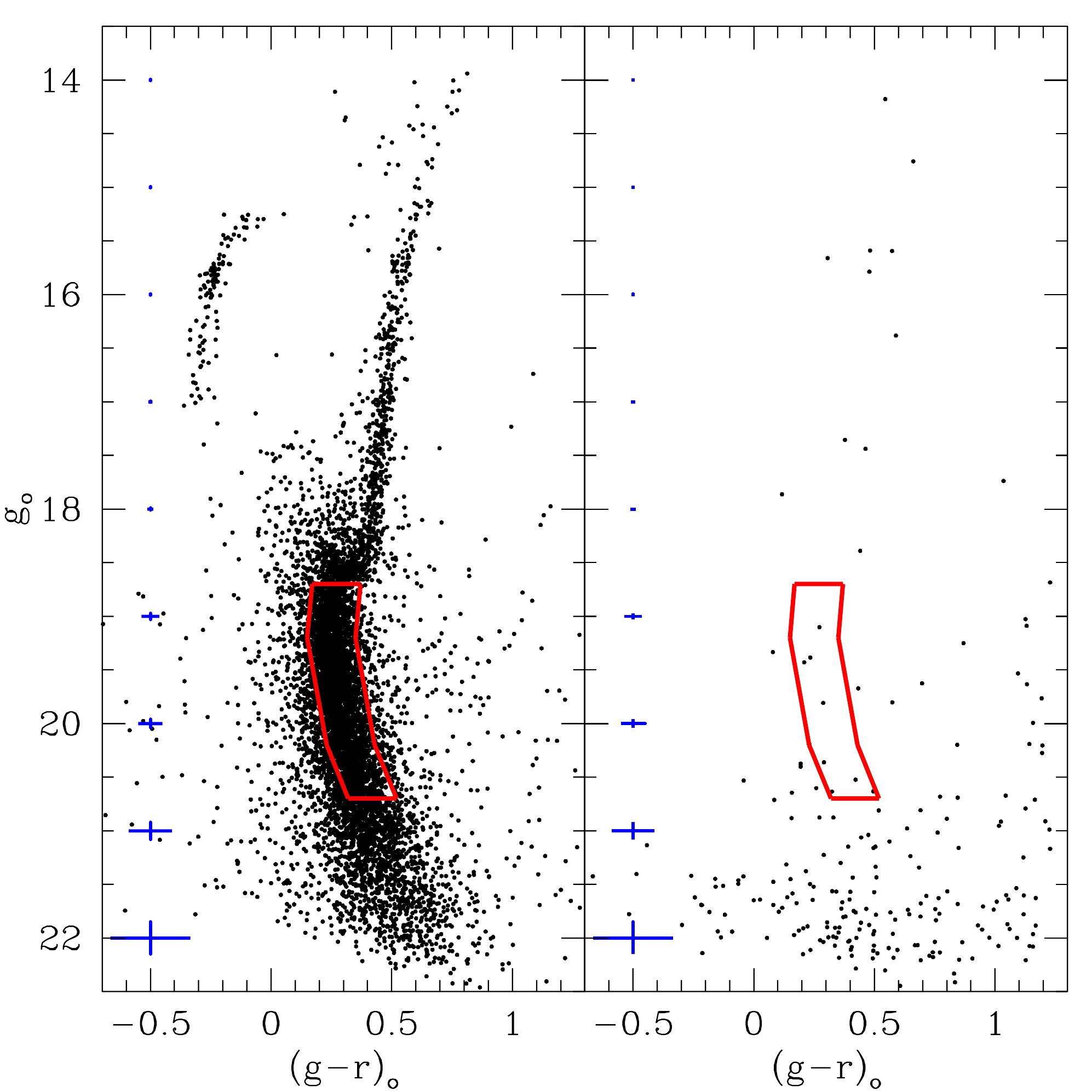}
    \caption{Colour-magnitude diagrams of stars in the field of NGC\,288 ($r$ $<$
5; left panel) and in a same area MW field located $\sim$ 1.5 degree towards to
north-east (right panel). Errorbars are included at the left margin of each
panel (blue lines). The region used
to perform star counts is overplotted with a red contour line.}
   \label{fig:fig1}
\end{figure}

In order to clearly identify and trace tidal tails in the field of NGC\,288,
we need to cover homogeneously a large sky area centred on the cluster with
deep photometry. For this purpose, we made use of the public
astrometric and photometric catalogue produced by
 the Panoramic Survey Telescope and Rapid response System 
\citep[Pan-STARRS PS1\footnote{The Pan-STARRS1 Surveys (PS1) and the PS1 public
science archive have been made possible through contributions by the Institute for
Astronomy, the University of Hawaii, the Pan-STARRS Project Office, the Max-Planck
Society and its participating institutes, the Max Planck Institute for Astronomy,
Heidelberg and the Max Planck Institute for Extraterrestrial Physics, Garching, 
The Johns Hopkins University, Durham University, the University of Edinburgh, 
the Queen's University Belfast, the Harvard-Smithsonian Center for Astrophysics, 
the Las Cumbres Observatory Global Telescope Network Incorporated, the National 
Central University of Taiwan, the Space Telescope Science Institute, the National
Aeronautics and Space Administration under Grant No. NNX08AR22G issued through the 
Planetary Science Division of the NASA Science Mission Directorate, the National
Science Foundation Grant No. AST-1238877, the University of Maryland, Eotvos Lorand 
University (ELTE), the Los Alamos National Laboratory, and the Gordon and Betty Moore 
Foundation.}
;][]{chambersetal2016}. We downloaded positions (R.A. and Dec.) and $gr$ PSF
photometry for 316984 stars distributed in a box of 4$\degr$$\times$4$\degr$ centred 
on NGC\,288. Aiming at illustrating this wealth of information,
Fig.~\ref{fig:fig1} depicts the intrinsic colour-magnitude diagram (CMD) of the inner 
cluster region ($r$ $<$ 5 arcmin (12.9 pc)) and that of a same area star field located $\sim$ 1.5
degree towards the north-east. To derive intrinsic magnitudes $g_o$ and colours
$(g-r)_o$ we corrected the Pan-STARRS PS1 $gr$ magnitudes by interstellar
absorption using the $E(B-V)$ values of each individual star obtained from
\citet{sf11}, which is the recalibrated Milky Way (MW) extinction map of  
\citet{schlegeletal1998}. The average colour excess for the whole surveyed field  is 
$E(B-V)$ = 0.014$\pm$0.001 mag. This means that the sky area used in this analysis
is affected  by a low colour excess with no sign of differential reddening.

\section{Star counts}

To trace the cluster stellar density radial profile and to build the stellar density map
of the analysed region, we considered stars along a strip of the
cluster main sequence (MS), from its main sequence turnoff (MSTO) down to 2 mags,
as is depicted with red lines in Fig.~\ref{fig:fig1}.  It expands $g_o$ 
magnitudes and $(g-r)_o$ colours in the range (18.7,20.7) and (0.15,0.52), respectively.
Notice that the strip was traced using stars within 5 arcmin (12.9 pc) of the NGC\,288's
centre, where more of them are probably cluster members. This assumption
is particularly supported by the fact that the same region in the CMD for
any MW field located well beyond the cluster body does not seem to
contain so many stars (see Fig.~\ref{fig:fig1}).
The selected  stars, because of their smaller masses, can be found far away 
from the cluster main body
\citep[][and references therein]{carballobelloetal2012}, and have been used
previously for searching extra-tidal structures in different GGCs 
\citep[see, e.g.][]{olszewskietal2009,sahaetal2010,p17c}. We decided to go as
deep as to be within 100 per cent of the photometry completeness; the
50 per cent photometry completeness being at $g$=$r$= 23.2 mag, determined with
PSF photometry of stellar 
sources in the stacked images (Farrow et al., in preparation).

\begin{figure*}
	\includegraphics[width=\columnwidth]{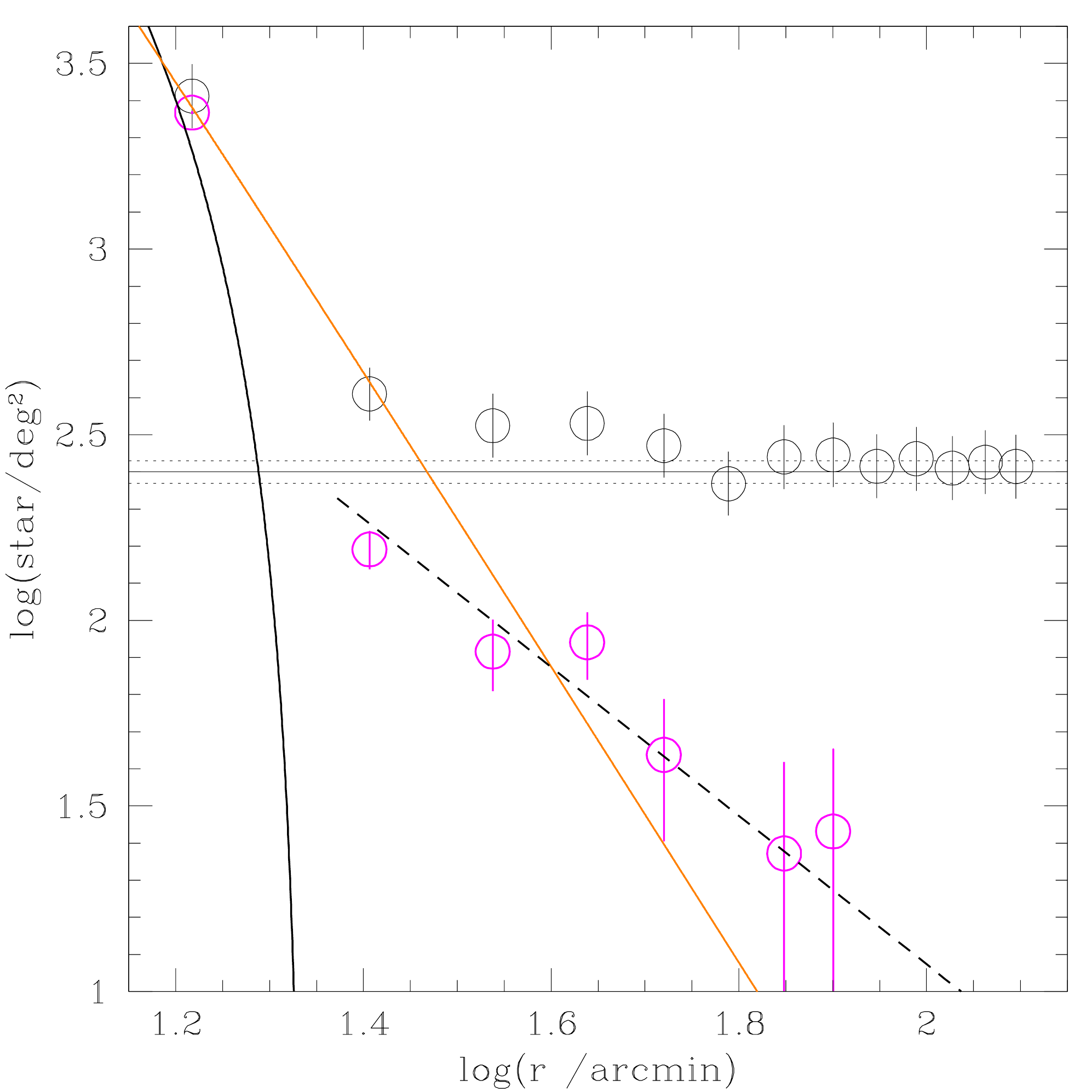}
	\includegraphics[width=\columnwidth]{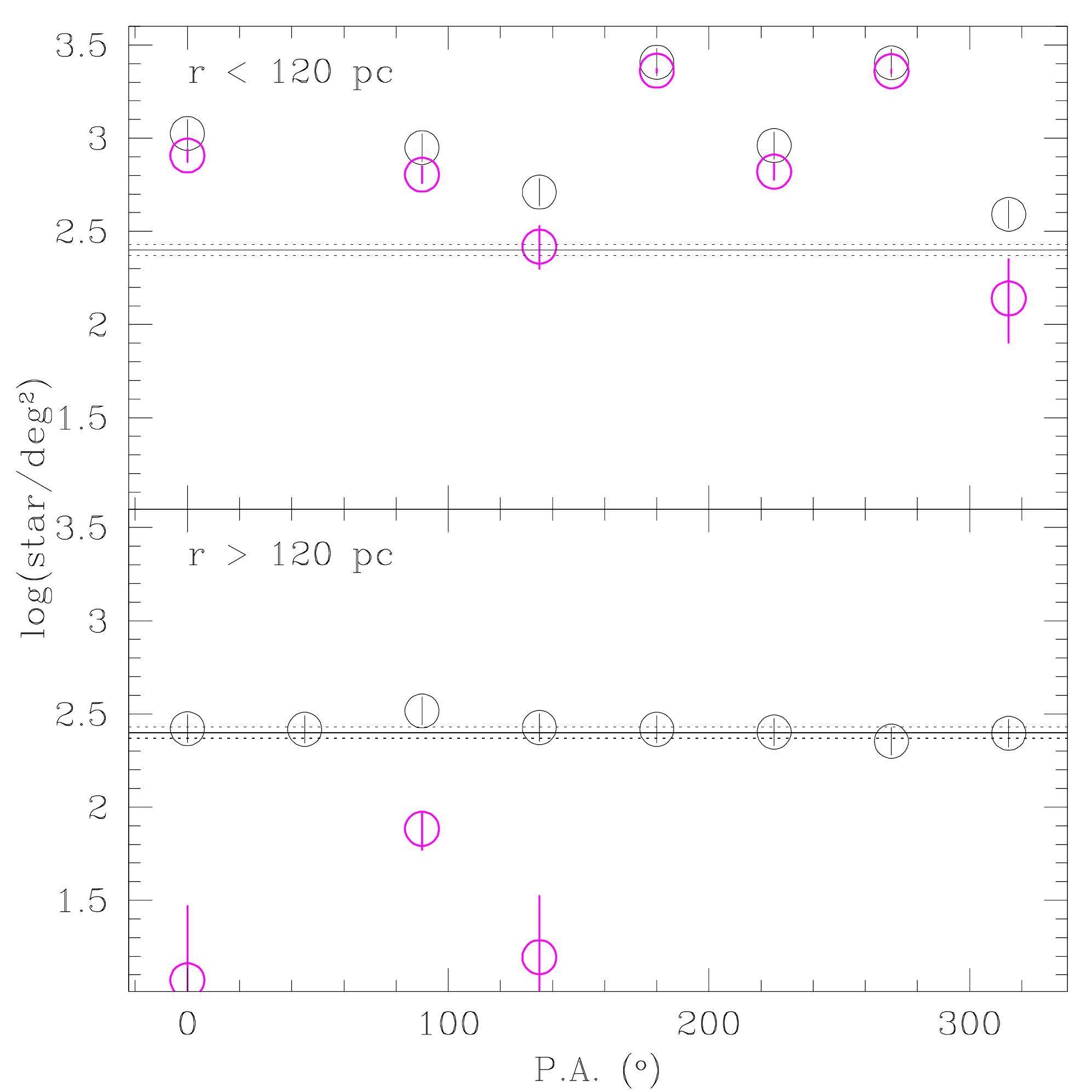}
    \caption{Left: Observed (black) and background subtracted (magenta) stellar
    density radial and profiles (open circles) with their respective errorbars. The 
    horizontal lines represent
    the mean background level and its dispersion, while the curved black
    and orange lines are the \citet{king62}'s and \citet{plummer11}'s models,
    respectively, with $r_c$, $r_h$ and $r_t$ taken from \citet{miocchietal13}.
    The dashed line represents a power law  $\propto$ $r^{-2}$.
    To convert angular to linear distances  we used a cluster distance of 8.9 kpc
    \citep{harris1996}. Right: same profiles as a function of the P.A. for
stars located closer and farther than 120 pc (0.8$\degr$) from the cluster centre.
    }
   \label{fig:fig2}
\end{figure*}

\begin{table}
\caption{Positions and covered areas of control fields. }
\label{tab:table1}
\begin{tabular}{@{}lccc}\hline
 Method  & Relative R.A.cos(Dec.)  & Relative Dec.   & size  \\
         &  (deg)  & (deg)  & (deg$^2$) \\\hline

Star     & -2.0 to 2.0 & 1.5 to 2.0 &  4.0$\times$0.5\\
counts   & -2.0 to -1.5 & -2.0 to 1.5 &  0.5$\times$3.5\\\hline
         &   -1.75           &    1.75         &   0.5$\times$0.5  \\
CMD      &   1.75       &    1.75         &   0.5$\times$0.5  \\
cleaning &   1.75     &   -1.75          &   0.5$\times$0.5 \\
         &     -1.75       &    -1.75         &  0.5$\times$0.5   \\\hline
 \end{tabular}
\end{table}

To count stars distributed throughout the surveyed region and within the 
defined MS strip,
we first split the total analysed field in small adjacent boxes of 
0.10$\degr$$\times$0.10$\degr$ that covered the entire area. Then, we counted
the number of MS strip stars inside them and computed the mean stellar
density as a function of radius by averaging the star counts in every box
placed within rings centred on the cluster of radius $r$ and $r$+$\Delta$($r$). 
Thus, we also estimated an uncertainty in stellar counts due to stellar
fluctuations in each ring. We repeated this exercise with boxes of increasing
size in steps of 0.01$\degr$ per side, up to 0.20$\degr$$\times$0.20$\degr$.
On the other hand, we also considered that a star can fall outside the MS
strip because of its photometric errors (see errorbars in Fig.~\ref{fig:fig1}), 
and thus change the total number
of stars within the MS strip. Notice that this is not an issue of
incompleteness in the photometry, since we decided to used stars brighter that the 
magnitude at the 100
per cent of photometry completeness.
We then averaged all the generated individual stellar density radial profiles, and the 
resultant one is depicted in Fig.~\ref{fig:fig2} (left panel) with black open circles. 
We have considered outer cluster regions
where the Pan-STARRS PS1 data set is at a 100 per cent completeness level.

The background level was estimated using a relatively large area embracing the
surveyed one (see Table~\ref{tab:table1}) from which we computed the mean and standard deviation values.
The resultant dispersion takes into account spatial variation of MW field stars
distribution in the MS strip. As can be seen in Fig.~\ref{fig:fig2}  the MW field
is quite uniform at that high Galactic latitude (b=-89.38$\degr$), as judged by the small
separation of the dotted horizontal lines to the solid line which
represent the dispersion and mean values, respectively.

We then subtracted such a MW background level from the observed radial profile
to derive that of NGC\,288, which we overplotted with magenta symbols. Here, the
errorbars come from considering in quadrature the MW field stellar density
dispersion and that of the observed radial profile. For comparison purposes
we superimposed the \citet{king62}'s and \citet{plummer11}'s models 
with black and orange curves, respectively, using core ($r_c$), half-light ($r_h$) 
and tidal ($r_t$) radii derived by \citet{miocchietal13}. An \citet{eff87}'s model
with the above $r_c$ value and $\gamma$ = 3.9 agrees pretty well with
that of \citet{plummer11}. As can be seen, NGC\,288
contains a population of extra-tidal stars reaching $\sim$ 3.5 times its tidal radius.

At this point, it is not possible the assess whether that extra-tidal structure
reveals the existence of tidal tails.  Indeed, some GGCs with a break in the radial
density profiles -- like the one observed in Fig~\ref{fig:fig2} (left panel) -- 
present tidal tails \citep[e.g., Eridanus and Pal\,15;][]{myeongetal2017}, while others
do not \citep[e.g., NGC\,1261 and NGC\,7089;][]{kuzmaetal2016,kuzmaetal2017}. In 
order to address this issue, we investigated
whether the resultant stellar density radial profile may vary with the 
orientation from the cluster centre. For that purpose, we repeated the above analysis distinguishing different position angles (PAs). We used angular sections of 45$\degr$ and performed two 
star counts, one for stars closer than 120 pc (0.8$\degr$) from the cluster centre,
and another for stars farther than that distance. We obtained the results shown in Fig.~\ref{fig:fig2} (right panel). As can be seen, while the upper panel shows a variable stellar density as a function of the P.A., the lower panel suggests that
the excess detected above the \citet{plummer11}'s profile (see left panel of Fig.~\ref{fig:fig2}) 
is related to an excess of stars located towards the east of NGC\,288 
(central P.A. $\sim$ 90$\degr$).
In the case of existing tidal tails, symmetrically collimated structures 
should have been detected well beyond the cluster body, which are not seen in
the figure.

\section{CMD cleaning}

Because we aim at disentangling very faint stellar structures in the
surrounding region of NGC\,288, the intrinsic stellar density map
of the cluster must be built on the basis of cluster members distributed
over the surveyed region. Fortunately,
the cluster is located at a high Galactic latitude (b=-89.38$\degr$)
where variation of the MW field are mostly negligible (see Section 3). 

In order to decontaminate the analysed area from the actual number of
MW field stars at any particular position, we decided to clean the MS strip
by statistically subtracting the MW stars that fall in that CMD region and that are
located far from the cluster region, but not too far as to lose the
local distribution in stellar density, magnitude and colour of MW stars.
Notice that this approach is different from that based on a spatial 
filter analysis, and is advantageous because we dealt with intrinsic 
cluster CMD feaures \citep{olszewskietal2009,p17c}.
The reference MW fields were chosen to be located to the north, east, south
and west from the cluster, at a distance of 2.5 degree and with areas of 
0.5$\degr$$\times$0.5$\degr$ each (see Table~\ref{tab:table1}). From these regions, we built four CMDs
and generated a sample of boxes ($g_o$,$(g-r)_o$) centred on each MW field
star, with sizes ($\Delta$$(g)$,$\Delta$$(g-r)$) defined in such a way that
one of their corners coincides with the closest MW field star in that CMD.
This procedure to map the MW field CMD was developed by \citet{pb12} and 
successfully used elsewhere \citep[see, e.g,][]{p17b,p17c,petal2017}. It has the 
advantage of accurately reproducing the reference star field in terms of
stellar density, luminosity function and colour distribution.

\begin{figure*}
	\includegraphics[width=\textwidth]{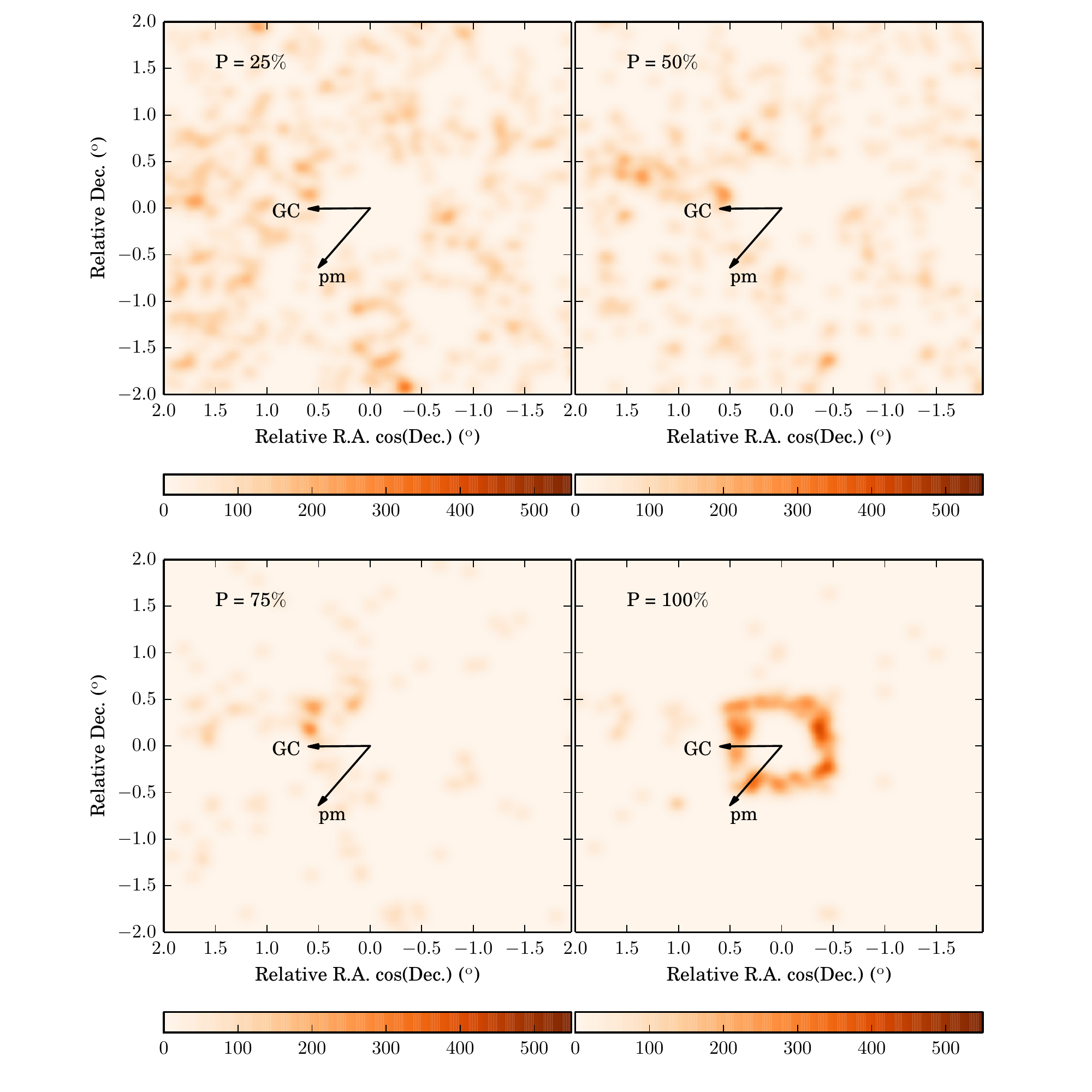}
    \caption{Stellar density maps with stars within different photometric
    membership probability ranges, as indicated at the top-left margin of each 
panel. The colourbars give the number of stars/deg$^2$. In order to highlight 
    the outer cluster regions, we cut off stars located inside $r_t$. The
 arrows represent the cluster absolute proper motion vector \citep[pm;][]{dinescuetal1999,leonetal2000}
 and that towards the Galactic centre (GC).
}
   \label{fig:fig3}
\end{figure*}

Each of these four generated CMD box samples were superimposed to the 
CMDs of regions of 0.5$\degr$$\times$0.5$\degr$ distributed throughout the 
studied area around NGC\,288 and subtracted from them one star per box; that
closest to the box centre. The resultant cleaned CMD contains mainly 
cluster members, although some negligible amount of interlopers can be expected.
Since we repeated this procedure four times, with reference MW fields 
strategically distributed, we could assign photometric
cluster membership by counting the number of times a star kept unsubtracted. Thus,
stars subtracted three times have photometric membership probabilities $P$ = 25\%,
and mainly represent field populations projected on the cluster area;
those subtracted two times, $P$ = 50\%, which could 
indistinguishably belong to the field or to the cluster; those subtracted once, $P$ = 75\%, 
i.e., stars that are predominantly found in the cluster rather than in the star field population, and
those kept unsubtracted, $P$ = 100\%.  We generated
stellar density maps of cluster members with $P$ = 25\%, 50\%, 75\% and
100\%, respectively.
Such  density maps are depicted in Fig.~\ref{fig:fig3}. 
Notice that stars with a statistical low probability of being cluster members ($P$ = 25\%)
do not trace any extended stellar structure around the cluster as that
 clearly visible in the  $P$ = 100\% panel. There are small group of stars spread throughout the
entire field.
Particularly, an stellar excess located at (Relative R.A., Relative Dec.) $\approx$ 
(1.5$\degr$, 0.5$\degr$) -- most clearly seen in the $P$ = 50\% panel --
 agrees with the excess observed in Fig.~\ref{fig:fig2}, namely, a stellar overdensity
that contributes to the extended cluster stellar radial profile at distances larger than
120 pc from the cluster centre (67.5$\degr$ $\le$ P.A. $\le$ 112.5$\degr$, 
centred at P.A. = 90$\degr$).

We smoothed the star distribution with a $\sigma$=3.6$\arcmin$ Gausian kernel, 
much smaller than the 16 arcmin  of spatial resolution used by 
\citet{leonetal2000}. 
The resultant density map for $P$ = 100\% illustrates how the cluster 
density diminishes with increasing 
distances from its centre, and that a clumpy pattern with different mean densities 
are seen
around it. It is in excellent agreement with the independent results found
of Section 3, i.e.,  NGC\,288 simply contains an extra-tidal structure; 
no long tidal tails oriented in the direction of the orbital motion 
 (see proper motion vector in Fig.~\ref{fig:fig3})
as those claimed by \citet{leonetal2000} and \citet{grillmairetal2004}
could be uncovered. However, \citet{montuorietal2007} 
and \citet{klimentowskietal2009} have argued from N-body simulations that tidal tails
near a cluster mainly point in the direction of the Galaxy centre (GC). For the sake of the
reader Fig.~\ref{fig:fig3} also illustrates the direction towards the GC. Seemingly,
it is not straightforward to link such a direction with the stellar excesses.
Nevertheless, some farther, few 
isolated small structures could be related to the cluster, if we considered stars with 
lower photometric membership probabilities.

\section{analysis and discussion}

Our both independent approaches lead to conclude that NGC\,288 is not a 
tidally limited GGC, but one with an extended structure that reaches
$\sim$ 3.5 times its tidal radius \citep{miocchietal13} and 
$\sim$ 2.5 times its Jacobi radius \citep[$r_J$ = 81.5 pc][]{baumgardtetal2010}. This could 
suggest that the cluster MS stars located in the outermost regions are
experiencing, in some way, gravitational effects due to the MW potential.
Tidal tails as those
claimed by \citet[][hereafter L00]{leonetal2000} and 
\citet[][hereafter G04]{grillmairetal2004} have not been
detected in this study. 

In order to find some explanation for the present negative outcome we
revisited the photographic CMD used by L00 (see their Figure 4).
That figure barely reaches tenth of magnitudes below the cluster MSTO, with clear
sign of photometry incompleteness (the darken the Hess diagram the more
numerous the observed cluster star population along the cluster MS). 
The error quoted by L00 at $B$ $\sim$ 20.0 mag is Pan-STARRS PS1 data set$\sigma$($B$) $\sim$ 0.2 mag, 
which is nearly
seven times larger than that in the Pan-STARRS PS1 data set used in the 
present work. On the other hand, L00 mentioned a not satisfactory background
subtraction, with residuals of Abell galaxies  that mimic cluster stellar
excesses beyond the cluster main body. We think that these three main
factors affected the final stellar density radial profile and stellar density map
built by L00, and that the astrometric and photometric data set used here
largely supersedes that previous photographic work. An additional difference
between L00 and the present results  is the derived slope $\alpha$ for a
power law  $\propto$ $r^{-\alpha}$
of the stellar density as a function of the distance to the cluster centre.
While L00 obtained a value of 1.18, we derived $\alpha$ = 2 (see dashed
line in Fig.~\ref{fig:fig2}). Our value is in between those found in GGCs 
with extended halo-like structures, e.g., NGC\,1851 and 47\,Tuc 
\citep[][$\alpha$ = 1.24]{olszewskietal2009,p17c} and the abrupt fall
of the $r^{-4}$ law suggested by \citet{penarrubiaetal2017}
as a prediction of expected stellar envelopes of GGCs embedded in dark mini-haloes.

G04's results are also hard to reproduce, namely, the existence of two main tidal tails
of $\sim$ 8.5$\degr$ long that arise from the cluster centre towards to the north-west
and south-east, respectively. One of both main tails have not been detected by L00, nor 
any of them in this work either. By revisiting the near-IR CMD used by G04 to claim such
a detection (see their Figure 1), we realized that that 2MASS photometry does not
reach the cluster MSTO, and that the CMD is largely more
contaminated by MW field stars than its counterpart in the optical regime 
(see Fig.~\ref{fig:fig1}). Moreover, the cluster red giant and horizontal branches are 
so far no visible in that $J$ versus $J-K$ CMD. Therefore, we are not aware of what
CMD cluster features were traced by G04.  All in all, both L00's photographic photometry and 
the Pan-STARRS PS1 data sets supersede that of 2MASS. Notice that GGCs with evidence of tidal
tails show a symmetric density pattern 
\citep[see, e.g.,][]{odenetal2001,belokurovetal2006,noetal2010,sollimaetal2011,balbinotetal2011,erkaletal2017,naverreteetal2017,myeongetal2017}, 
not distinguished in Fig.~\ref{fig:fig3}, either.

Our observational evidence about the extended structure of NGC\,288 does not seem to be fully 
in the line of those  argued  from a theoretical point of view 
\citep{go1997,dinescuetal1999,bg2017}, in the sense that disruption by tidal shocks were more 
important than internal relaxation and evaporation. \citet{dinescuetal1999}  listed 
eight GGCs (NGC\,288, 5139, 6121, 6144, 6362, 6712, 6769 and Pal\,5) with similar dynamics
supporting the existence of tidal tails; only Pal\, 5 having been confirmed. According to
recent works, NGC\,5139 \citep{fetal2015}, NGC\,6121 \citep{wvdm2017}, NGC\,4166, 6362, 6712 and 6779
\citep{chch2010} do not show evidence of tidal tails. NGC\,288, studied here, must now be included 
in this list, since the present outcomes suggest that Galactic tidal interaction has been a relatively 
inefficient process for stripping stars off the cluster. 
At this point it would be interesting to derive accurate GGC proper 
motions, for instance, from the next Gaia Data Release 2 (DR2) and to compute again their orbital motions and, 
on the other hand, to derive  stellar radial profiles in an homogeneous scale. Thus, it would be possible to 
search  for any relationship between extended structural features and orbital motions as to infer
whether GGC masses, or the number of passages near the Galactic centre, or any other orbital
parameter (e.g., eccentricity, inclination), or their birthplaces  have to deal with the structural 
features seen far away the clusters' centres.

\section{Conclusions}

The issue about the extended structures of GGCs has fueled a renewed debate with the advent
of large photometric surveys that allow us to cover homogeneously wider areas and 
analysed them from deep photometry. GGCs show a wide
variety of extended structures, from those having no signature for such a feature
to those with long tidal tails, passing through those exhibiting extra-tidal stellar
populations more or less extended, azimuthally distributed or as clumpy features.
Up-to-date, there is not a clear consensus for their origins and contradict results have been published for some of them.

Here we performed a sound analysis of the external region around NGC\,288, claimed by
L00 and G04 to have visible tidal tails, and supported by studies of its orbital
motion as a very good candidate to have long tidal tails. For this purpose, we
took advantage of the Pan-STARRS PS1 data set for an area of 4$\degr$$\times$4$\degr$ around the cluster.
From the cluster CMD we defined
a strip along the cluster MS where we carried out stars count in order to construct the  cluster stellar density radial profile and a stellar density map. 
 
The MW subtracted stellar density radial profile shows an extra-tidal population of cluster stars that extends up to $\sim$ 3.5 times
the cluster tidal radius. This is a moderate extended structure, since other GGCs show
evidence of such a features up to nearly more than 6 times their tidal radii (e.g., NGC\,1851,
47\,Tuc). 
The stellar density map built with stars that
have photometric membership probabilities equal or higher than 50 per cent reveals a
somehow clumpy structure around the cluster with different stellar densities, in 
excellent agreement with the resultant radial profile. The detected extra-tidal component
is well matched by a power law with $\alpha$ = 2. None of both independent approaches
shed light on the possibility of the existence of tidal tails.
This points to the
need of  more reliable orbital motions in order to constrain whether 
the number of passages near the Galactic centre, the eccentricity, the birthplaces, the masses,
among other parameters are responsible for the wide variety of extended GGC's features seen until
the present.


\section*{Acknowledgements}
We thank C. Grillmair and D. Cassetti-Dinescu for reading the manuscript and making fruitful comments, 
and the referee for the thorough reading of the manuscript and
timely suggestions to improve it.




\bibliographystyle{mnras}

\input{paper.bbl}







\bsp	
\label{lastpage}
\end{document}